\documentstyle[emulateapj,epsf]{article}

\newcommand {\simlt}{\lower.5ex\hbox{$\; \buildrel < \over \sim \;$}}
\newcommand {\simgt}{\lower.5ex\hbox{$\; \buildrel > \over \sim \;$}}

\newcommand {\lsim}{\lower.5ex\hbox{$\; \buildrel < \over \sim \;$}}
\newcommand {\gsim}{\lower.5ex\hbox{$\; \buildrel > \over \sim \;$}}

\begin{document}

\title{CMB Constraints on a Baryonic Dark Matter-Dominated Universe}

\author{Louise M Griffiths,
Alessandro Melchiorri, Joseph Silk}

\affil{ 
Nuclear and Astrophysics Laboratory, University of Oxford, \\
Keble Road, Oxford, OX 3RH, UK\\}

\begin{abstract}

In recent years, upper limits on the cosmic microwave 
background (CMB) anisotropies
 combined with predictions made by theories of galaxy formation,
have been extremely powerful in ruling out purely baryonic dark matter (BDM) 
universes.
However, it has recently been argued that the absence of a prominent second peak in the anisotropy
spectrum measured by the BOOMERanG-98 and MAXIMA-1 experiments
seems to favour a $\Lambda$BDM model when combined with standard
Big Bang Nucleosynthesis (BBN) constraints.
In this {\it Letter}, we further investigate this result
showing that, using the CMB data {\it alone}, 
a purely baryonic adiabatic model of structure formation
seems unlikely if the universe is flat ($\Omega=1$).
Combining the CMB data with supernova type Ia (SNIa) data renders 
purely baryonic models inconsistent with flatness at high
significance and  more than $3 \sigma$ away from both the 
BBN constraints and the HST key project result $h=0.72 \pm 0.08$.
These results indicate that only a radical revision of cosmology with
{\it ad hoc} properties could bring baryonic models
such as those advocated by MOND enthusiasts back 
into agreement with current observations.
\end{abstract}


\section{Introduction}

Since the pioneering work of Zwicky (1933), astronomers have found 
compelling evidence to suggest that the major contribution to the 
overall mass density of the universe is in 
the form of {\it dark} matter. Even if its precise nature remains 
unknown (see e.g. \cite{ber} for a recent review), the currently favoured 
hypothesis of {\it cold} dark matter (CDM) is in good
agreement with an enormous body of data: anisotropies in the cosmic 
microwave background (CMB), large-scale structure surveys, cluster 
abundances, structure of the Lyman-$\alpha$
forest and a host of other measurements (see e.g. \cite{bah}).
Furthermore, in the past 30 years, upper limits on the CMB anisotropies 
and predictions made by theories of galaxy formation have been
extremely effective  at ruling out 
models based on purely {\it baryonic} or {\it hot} 
dark matter (see e.g. \cite{dor}; \cite{wil}; \cite{kai}; \cite{whi}). 

The quality of astrophysical data has improved and it currently
seems  likely that the present fractional 
overall energy density $\Omega$ includes contributions 
from a cosmological constant $\Lambda$ ($\Omega_{\Lambda}$) and from cold 
($\Omega_{cdm}$), hot ($\Omega_{hdm}$) and baryonic ($\Omega_b$) dark 
matter.  Also, it appears that the densities of these components
are all within one or two orders of magnitude of each other.
While the similar densities of dark `energy' and `matter' can 
possibly be explained by advocating a quintessence scalar field
which `tracks' the matter density (\cite{ste}),
the reason why the densities of the $3$ matter components are so similar
is not so apparent.  

Predictions made by structure evolution 
theories based on the $\Lambda$CDM model disagree
with some galaxy observations,
which suggests that, for the model to hold, 
assumptions made about the properties of CDM particles must be modified
(see e.g. \cite{spe}; \cite{bur}; \cite{mad}; \cite{moo}).  If this
reasoning is adopted,  
currently discussed versions of the present cosmological scenario, incorporating
a fine-tuned cosmological constant and dark matter particles
with {\it ad hoc} properties,  lose the compelling aesthetic
simplicity of the original CDM model (see \cite{sellwood} for a 
recent review).  Of course, other, more astrophysical, resolutions
may be possible \cite{bin}. 
%(Binney, Gerhard and Silk 2001)
 
 The recent CMB anisotropy measurements from the BOOMERanG-98 
(\cite{debe}) and MAXIMA-1 (\cite{max}) experiments have provided new insights
into the cosmological parameters.
For the first time, the anisotropy angular power spectrum has been 
measured over a wide range of angular scales, from multipole $l \sim 50$ 
up to $l \sim 800$ with errors of the order of $10 \%$. 
The data sets confirm that there is a peak in the angular power 
spectrum at $l\sim 200$ with a steep decline 
in power from $l \sim 200$ to $l \sim 300$.
While the presence of such a peak is
consistent with the predicted acoustic oscillations in the 
adiabatic inflationary scenario, the absence of prominent 
secondary peaks after $\ell \sim 300$ 
is an unexpected result, suggesting that the value for the physical baryonic
density is $\sim 50 \%$ higher than that predicted by 
Big Bang Nucleosynthesis (BBN) (see e.g. \cite{tegz}; \cite{jaffe}).
We note however that this discrepancy with the standard model is no more than 2 $\sigma$
when the most recent CMB anisotropy results are included \cite{cbi}.

In recent months, several solutions have been advocated
to reconcile BBN with the larger values of the baryon density
suggested by the CMB data. In a recent letter by McGaugh 
(2000), the simple and controversial solution of 
considering only baryonic 
dark matter (BDM) plus a cosmological constant has been proposed.
The lower CMB second peak can thereby be explained by the collisional 
damping expected in a BDM
universe, while maintaining the baryon density at a level compatible
with the BBN constraint.
As is well known, purely baryonic adiabatic models are in severe 
contradiction with the measured amplitude and shape of galaxy and cluster 
correlations.
However, there remain serious issues about the possibly scale-dependent bias 
parameter. Hence McGaugh's result is interesting for various reasons:
it resolves the disagreement with BBN by using a simpler
model ($\Omega_{cdm}=\Omega_{hdm}=0$), it demonstrates the model-dependence
of theoretical fits to the
CMB data by exploring a previously uncovered corner of parameter
space and, finally, it questions our theoretical understanding
of the structure formation process.

If it turns out that the CDM hypothesis does
not work well at some cosmological epochs and scales, 
then even the exotic scenario of a purely baryonic model with 
modifications to standard gravity 
could become competitive, especially in view of the 
insistence of MOND proponents that this alternative theory of gravity
merits serious examination (\cite{mcg98}; \cite{san2k}). 
Thus, it is timely to study how compatible the
CMB data is with a BDM adiabatic universe, what consequences this assumption 
can have on the remaining cosmological parameters, and 
the need for CDM from independent observations of
galaxy clustering.

In this {\it Letter}, we perform an analysis 
in cosmological parameter space.
Using the recent CMB data {\it alone} we show 
that a purely baryonic adiabatic model appears unlikely
if the universe is flat ($\Omega=1$).
By further combining these results with supernova type Ia (SNIa) data,
we rule out baryonic adiabatic models 
with high confidence.

\section{Method}

The structure of the $C_{\ell}$ spectrum depends essentially
on 3 cosmological parameters (see e.g. \cite {huss};
\cite{efs}; \cite{mel} and references therein): the physical baryonic 
density $\omega_b=\Omega_bh^2$ and the overall matter density 
$\omega_m=\Omega_mh^2=(\Omega_{cdm}+\Omega_b)h^2$ which define the 
size of the acoustic horizon at decoupling,
and the `shift' parameter $R$ related to the geometry of the universe
through $R=2 \sqrt{|\Omega_k| / \Omega_m} / \chi(y)$,
with 
$y=\sqrt{|\Omega_k|}\int_0^{z_{dec}}
{[\Omega_m(1+z)^3+\Omega_k(1+z)^2+\Omega_{\Lambda}]^{-1/2} dz}$,
where the function $\chi(y)$ depends on the curvature of the universe 
and is $y$, $\sin(y)$ or $\sinh(y)$ for flat, closed and 
open models respectively.

A decrease in $\omega_{cdm}$, with $\omega_b$ and $R$ remaining constant, 
will reduce $\omega_m$. This 
decreases the redshift of equality causing the peak in the spectrum to be 
shifted towards smaller angular scales.
Constant $\omega_m$ can be maintained by increasing $\omega_b$. 
However, an increase in $\omega_b$ will
decrease the sound speed at decoupling 
$c_s\sim1/\sqrt{3(1+3\omega_b/(4\omega_{rad}(1+z_{dec})))}$
again reducing the acoustic horizon size and shifting the 
peak to higher $\ell$'s.

To keep the position of the first peak fixed
while $\omega_b$ and $\omega_m$ are varied, $R$ has
to be appropriately tuned.
This can be achieved by increasing  $\Omega_{\Lambda}$.
The shift parameter $R$ is sensitive to variations in 
$\Omega_{\Lambda}$ as $\Omega_m \rightarrow 0$, and for
$\Omega_m=\Omega_b \sim 0.05$ a $10 \%$ increase in 
$\Omega_{\Lambda}$ can produce a $50 \%$ decrease in
$\ell_{peak}$.
Thus, a purely baryonic model can produce the observed peak
structure at $\ell \sim 200$ only if $R$ is increased with respect 
to the corresponding flat 
CDM model. We can therefore anticipate that viable purely baryonic models
will be $\Lambda$-dominated with closed geometries.

Motivated by these considerations, we compare 
recent CMB and SNIa observations with a set of models with 
parameters sampled as follows: 
$\Omega_{m}= \Omega_{b} = 0.015, ...,0.5$; 
$\Omega_{\Lambda}=0.80, ..., 1.04$ and $h=0.40, ..., 0.95$.
We vary the spectral index of the primordial density perturbations
within the range $n_s=0.50, ..., 1.50$ 
and we rescale the fluctuation amplitude by a
pre-factor $C_{10}$, in units of $C_{10}^{COBE}$.
The theoretical models are computed using the publicly available 
{\sc cmbfast} program (\cite{sz}) and are compared with the recent 
BOOMERanG-98 and MAXIMA-1 results.
The power spectra from these experiments were estimated in 
$12$ and $10$ bins respectively, spanning the range
$25 \le \ell \le 785$. In each bin, the spectrum is assigned
a flat shape, $\ell(\ell+1)C_{\ell}/2\pi=C_B$.

Following Bond, Jaffe \& Knox (1999) we approximate the signal $C_B$ inside
the bin to be an offset lognormal distribution, such that the
quantity $D_{B}={\rm ln}(C_{B}+x_B)$ (where $x_B$ is the
offset correction) is a Gaussian variable. 
The likelihood for a given cosmological model is then
 defined by 
$-2{\rm ln} L=(D_B^{th}-D_B^{ex})M_{BB'}(D_{B'}^{th}-D_{B'}^{ex})$
where $C_B^{th}$ ($C_B^{ex}$) is the theoretical (experimental)
band power, $x_B$ is the offset correction and $M_{BB'}$ is
the Gaussian curvature of the likelihood matrix at the peak. 
 We consider $10 \%$ and $4 \%$ Gaussian distributed 
calibration errors for the BOOMERanG-98 and 
MAXIMA-1 experiments respectively. 
We also include the COBE data using Lloyd Knox's RADPack packages.
Proceeding as in Dodelson \& Knox (2000), we attribute a likelihood to a point
in the ($\Omega_m=\Omega_b$, $\Omega_\Lambda$) and
($\Omega_b$, $h$) planes by finding the remaining $3$ parameters that
maximise it. We then define our $68\%$, $95\%$
 and  $99\%$ contours to be where the likelihood falls to $0.32$, $0.05$ and
$0.01$ of its peak value, as would be the case for a
two dimensional multivariate Gaussian.

\section{Results}
In the top panel of Figure 1, we plot likelihood contours in the  
($\Omega_m=\Omega_b$, $\Omega_\Lambda$) plane by applying the 
maximization/marginalization algorithm described above and
using only the BOOMERanG-98 data and limiting our analysis 
to models with age $t_0 > 10 G{\rm yr}$.
As expected, the low sound speed at decoupling due to
the high baryon content makes only closed models compatible 
with the observations.
The deviation from flatness becomes less and less important 
as $\Omega_{b}\rightarrow0$, as one would expect from the expression for $R$. 
However, the decrease in the redshift at 
equality causes positive curvature models to be preferred.

In the parameter range we are sampling, we find that flat
models are excluded at $\sim 2 \sigma$.
In order to further test the discrepancy with a
flat universe we also include CDM models in the analysis
with $\Omega_{cdm}=0.05,...,0.45$. Comparing our models to the BOOMERanG-98
data alone and restricting the analysis to $\Omega=1$, we find that
$h^2\Omega_{cdm} > 0.04$ at the $95 \%$ C.L. Thus flat BDM models 
are excluded with significance.

For the BOOMERanG-98 analysis with $\Omega_{cdm}=0$, 
the best-fitting model is a closed model with
 $\Omega_{B}=0.135$, $\Omega_{\Lambda} =1.0$,
$n_{S}=0.94$, $h = 0.45$, $C_{10}=0.5$ and the
BOOMERanG-98 calibration left untouched.
This provides a good fit to the data (see Figure 2, top panel).
Including the MAXIMA-1 data points produces a different best-fitting model:
$\Omega_{B}=0.035$, $\Omega_{\Lambda} =1.00$,
$n_{S}=0.92$, $h = 0.85$, $C_{10}=0.5$ with a $\sim 12 \%$
upward calibration 
%\medskip
{\centerline{\vbox{\epsfxsize=8.0cm\epsfbox{fig1.eps}}}
{\small F{\scriptsize IG}.~1.--- 
The likelihood contours in the ($\Omega_B$, $\Omega_\Lambda$) plane,
with the remaining parameters taking their best-fitting values.
Only the BOOMERanG-98 data is considered here. The contours correspond to  
0.32, 0.05 and 0.01 of the peak value of the likelihood, which are the 68\%, 95\% and 99\% confidence levels respectively. 
The dashed line going from left to right diagonally across the plot indicates 
flat models.
The top panel is with a prior on the
age of the universe $t>10$ Gyr. 
The bottom panel includes the BBN prior 
$\Omega_bh^2 =0.019 \pm 0.002$ and the SNIa data.
The results of these two priors are incompatible at $\sim 3 \sigma$.
\label{fig:like_mar1}}
}
\medskip

\noindent
for the BOOMERanG-98 points and a $\sim 5 \%$ downward calibration for MAXIMA-1.

As the low value of $C_{10}$ already suggests, these best-fitting
models are in contradiction with the COBE data.
Including COBE in the BOOMERanG-98 analysis does 
not significantly affect the best-fitting cosmological parameters which remain
$\Omega_{B}=0.135$, $\Omega_{\Lambda} =1.0$,
$n_{S}=0.94$, $h = 0.45$, but the model now requires
$C_{10}=0.8$ and an upward shift in the calibration for BOOMERanG-98
of $\sim 30 \%$.  This model is ruled out by the COBE and BOOMERanG-98
data at the 95\% C.L., although this discrepancy can be resolved
by adding a gravitational wave (GW) contribution on large scales.
The top panel of Figure 2 indicates that the inclusion
of a GW component such that $C_{10}^T=C_{10}^S$
leaves the small scale behaviour practically unchanged
but provides enough CMB power on large scales to
match the COBE normalization.

A GW contribution as large as $C_2^T/C_2^S\sim 1$
is incompatible at the $95\%$ C.L. 
with flat $\Lambda$CDM models
when the BBN prior is assumed in CMB analyses (\cite{kmr}).
GWs leave a characteristic imprint on
polarization power spectra, via fluctuations in
the $B$ magnetic-type-parities channel (\cite{kam},
\cite{sel96}) that vanish in the case of scalar fluctuations.
However, the expected polarization amplitude for BDM
models is very low, below $0.5 \mu K$, and future experiments 
will not have the sensitivity to detect the
$B$ channel signal unless a substantial amount of
reionization occurred in the past (\cite{sel96}).

%{\centerline{\vbox{\epsfxsize=8.5cm\epsfbox{bestfit.ps}}}
{\centerline{\vbox{\epsfxsize=7.5cm\epsfbox{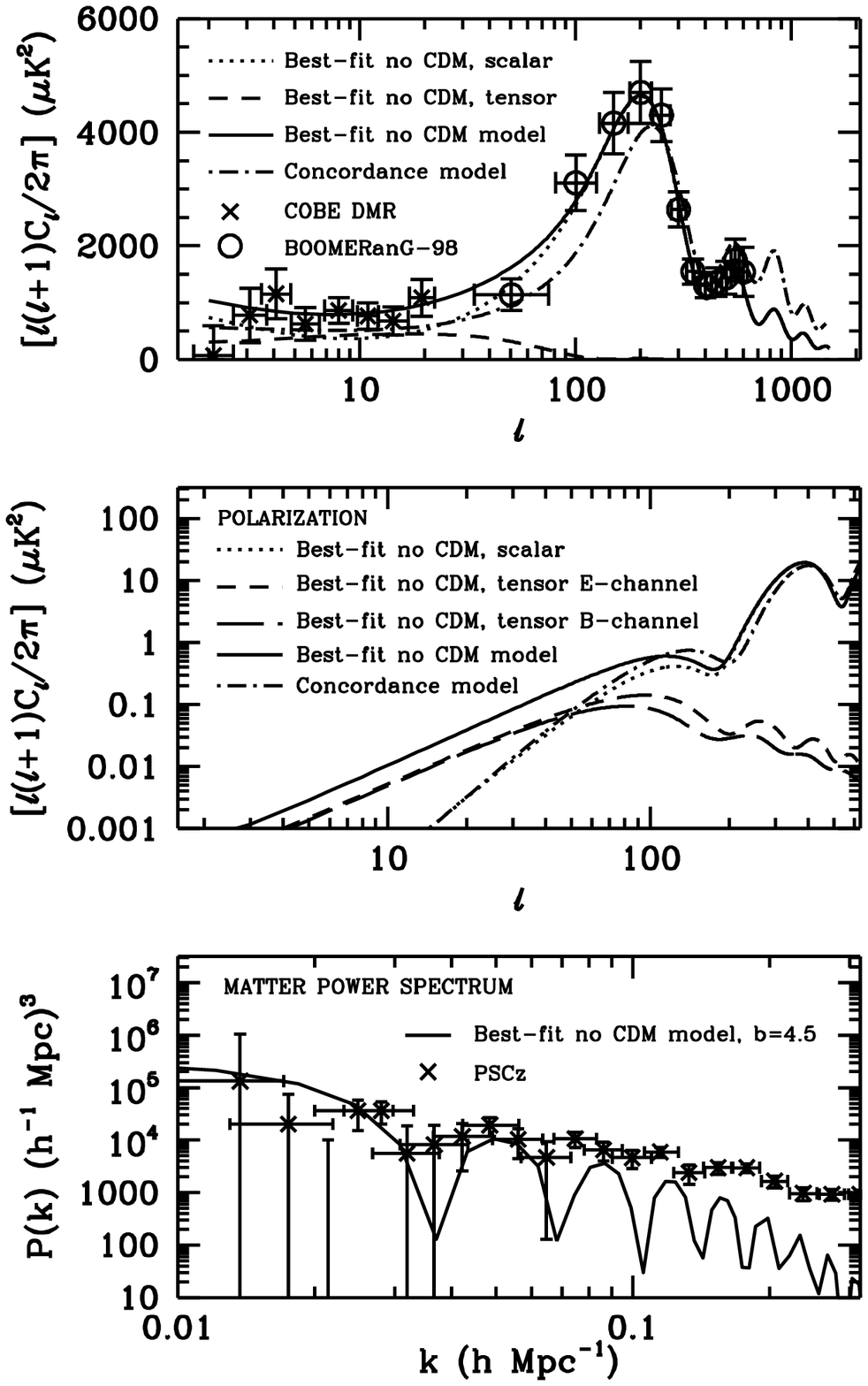}}}
{\small F{\scriptsize IG}.~2.--- 
In the top panel, the BOOMERanG-98 and the COBE data are plotted together 
with the best-fitting baryonic model to the BOOMERanG-98 data. 
In order to fit the COBE data, a tensor component must be added. 
For comparison, the $\Lambda$-CDM
concordance model is also plotted.
The middle panel shows the expected polarization signal of 
the best-fitting model. Including the tensor components produces $B$ 
modes in the polarization spectra.
In the bottom panel it can be seen that the best-fitting model 
to the BOOMERanG-98 data is in contradiction with 
the matter power spectrum measured by the PSCz survey.
\label{fig:like_mar}}}
\medskip

One can further combine the CMB constraints with 
those obtained from the luminosity--distance measurements of
high-$z$ supernovae (\cite{SN1a,SN1b}) as we do in the
bottom panel of Figure 1. The inclusion of the SNIa data
does not greatly affect the constraints on $\Omega_{\Lambda}$ but,
as expected, it rules out any models with low values of $\Omega_{b}$.
We find that $\Omega_b \ge 0.13$ at the $95 \%$ C.L.
 and an increasing inconsistency of BDM models with a
flat universe.

Also plotted in the bottom panel of Figure 1 are the
constraints obtained including a prior 
$\Omega_bh^2=0.019 \pm 0.002$ (\cite{Burles}) 
inferred by measurements of primordial elements and assuming standard BBN.
The results of the $2$ priors
are incompatible at more than $3 \sigma$.

It is interesting to further explore the compatibility of
our results with independent observations
by considering likelihood contours in the ($\Omega_{b}$, $h$)
plane.  Figure 3 plots the results of our analysis with no priors,
the results including the SNIa data and the results 

\medskip
%\centerline{\vbox{\epsfxsize=11.5cm\epsfbox{panel_omh.ps}}}
\centerline{\vbox{\epsfxsize=7.5cm\epsfbox{fig3.eps}}}
{\small F{\scriptsize IG}.~3.--- 
The likelihood contours in the ($\Omega_{B}$, $h$),
with the remaining parameters taking their best-fitting values
to the BOOMERanG-98 data. The contours correspond to  
0.32, 0.05 and 0.01 of the peak value of the
likelihood, which are the 68\%, 95\% and 99\% C.L. respectively. 
The filled contours correspond to an analysis
with a $t>10$ Gyr prior. The dashed, dotted and solid contours 
to the right of the plot are the
result when the SNIa data is included.
The bold black line to the left of the plot is the $95 \%$ C.L. 
region from BBN.
\label{fig:like_omh}}\medskip
\noindent
with the BBN constraint.
We can see from this figure that purely baryonic models are in quite good
agreement with BOOMERanG-98 and with the BBN constraints. However, when the
SNIa data is included we are restricted to models with
values of $\Omega_b$
that are too high and values of $h$ that are too low
to be consistent at $<3 \sigma$ with both the BBN constraint
and the recent HST result $h=0.72 \pm 0.08$ (\cite{free}).

Quite independently of the prior assumed, we find
that the age of the universe in this scenario is
constrained to be $t_0=24 \pm 2$ Gyr, nearly double
that expected from standard CDM.
The age of the universe must exceed the
ages of the oldest globular clusters $t_{GC}=14 \pm 2$ Gyr by 
an amount $\Delta t \sim 0.5 - 2$ Gyr, so the age of the universe
required by these models seems to be too high
unless globular cluster formation is delayed for a very long time.

Collisional damping erases
fluctuations in the matter power spectrum on scales
$< 100 h^{-1} Mpc$. 
The predicted matter power spectrum from the best-fitting model 
(see Figure 2, bottom panel) 
is in contradiction with the decorrelated linear power 
spectrum extracted from the recent PSCz catalogue (\cite{ham}).  
Allowing for a shift in the overall amplitude
with a bias factor $b$ where $P(k)_{PSCz}=b^2P(k)$, 
we obtain a best-fitting bias $b = 4.5$ with
a $\chi^2 = 155$ which is an extremely poor fit to 22 data points 
with one free parameter. The theoretical variance 
in baryonic matter fluctuations over a sphere
of size $8 h^{-1} {\rm Mpc}$, for example, is $\sigma_8 \sim 10^{-2}$ 
to be compared with the observational value 
$\sigma_8= 0.56\Omega_m^{-0.47}\sim1.4$
(\cite{liddle}). So, rescaling the galaxy
data with a massive bias parameter may not be physically meaningful since
no galaxies would have formed in order to allow there to be a
predicted galaxy power spectrum. 

Various phenomenological mechanisms can be proposed in an attempt to
resolve the discrepancy with the matter power spectrum and render
BDM models viable.  One such possibility is to increase
the spectral index of the primordial density fluctuations, $n_s$.
However, the 'blue' models that are able to produce $\sigma_8 \sim
0.5$ require $n_s \geq 2$ which is outside the range
$ 0.7 \lsim n_s \gsim 1.2$ allowed by inflation (\cite{stenr}).  Also, such a high tilt will cause the CMB spectrum to become incompatible with observations.

Another option that could increase small scale density perturbations whilst leaving the CMB spectrum unchanged is to have a spectral index that varies with scale, ${\rm d}n_s/{\rm d} \ln k \neq 0$, as predicted by some inflationary models (\cite{kos}).  Unfortunately, the variation needed, $n_s \gsim 1$, is much greater than that predicted by the most viable models.

More recently, there has been some investigation into features in the
primordial power spectrum generated during inflation (\cite{grif}; \cite{bar}; \cite{kne}).  It is possible that such features could
increase the perturbations on small scales without affecting the CMB
spectrum, although it is doubtful that a single feature would be able to resolve the observed discrepancy and such a mechanism is unlikely to be preferable to the assumption of CDM itself. 

\section{Discussion}
We have examined CMB anisotropies and large-scale structure observations
in a purely baryonic dark matter universe.  Our results suggest 
that a BDM adiabatic universe with a
power-law primordial power spectrum can only reproduce
the sub-degree CMB measurements if the universe is closed.
Although the premise that $\Omega=1$ is considered to be one
of the basic predictions of the inflationary scenario, it is possible
to construct inflationary scenarios that predict closed universes 
(see e.g. \cite{lin}; \cite{sta}).
It is interesting to note, however, that removing 
CDM immediately forces us to construct a 
more elaborate inflationary model, or, in other words,
the most simple model needs non-baryonic matter to work.

When flatness was assumed, we found that $\omega_{cdm} >0.04$
at $2 \sigma$ with $L(\omega_{cdm}=0)/L(\omega_{cdm}=0.1) \sim 10^{-3}$.
Therefore, the flatness constraint with BBN renders 
$\Lambda BDM$ models less consistent with the CMB observations than the standard $\Lambda CDM$ model.
Also, $\Lambda BDM$ models that fit the BBN constraints 
are degenerate with a non-physical region of the parameter space
(with very high baryon content and small $h$). When any 
cosmological prior other than BBN is assumed in the analysis, the unphysical region
turns out to be preferred.
Including the SNIa constraints (or $\Omega_m=\Omega_b > 0.25$), 
for example, makes the parameter space 
incompatible at $> 95 \%$ confidence with both the HST 
constraint $h=0.72 \pm 0.08$ and with BBN.

As expected, all the BDM adiabatic models fail to reproduce both the 
CMB observations and the observed amount of galaxy clustering 
on $8 h^{-1} {\rm Mpc}$ scales.
In order to solve all the discrepancies, that are not present 
in the standard $\Lambda$CDM scenario, we need to introduce 
{\it ad hoc} mechanisms which are unlikely to be 
preferable to the assumption of CDM itself.
Nonetheless, BDM models leave a set of characteristic
imprints such as large scale $B$-mode polarization and no third acoustic peak
in the anisotropy spectrum that will allow future experiments to 
further scrutinize this hypothesis.

\acknowledgements 
We thank Pedro Ferreira, Arthur Kosowsky, Andrew Liddle, 
Francesco Melchiorri, Gary Steigman for useful conversations and
the High-Z Supernova Search Team
for providing the SNIa likelihoods.

\end{document}